%% file: article.tex
\documentclass{svproc}
\usepackage{amsmath,amssymb,amsfonts}

\usepackage{url}

\begin{document}
\mainmatter              
\title{wsGAT: Weighted and Signed Graph Attention Networks for Link Prediction}
\titlerunning{Weighted and Signed Graph Attention Networks}  
%
\author{Marco Grassia\inst{1} \and Giuseppe Mangioni\inst{1}}
\authorrunning{Marco Grassia et al.} 
%
\tocauthor{Ivar Ekeland, Roger Temam, Jeffrey Dean, David Grove,
Craig Chambers, Kim B. Bruce, and Elisa Bertino}
\institute{Department
of Electric Electronic and Computer Engineering, University of Catania, Catania 95100, ITALY\\
\email{\{marco.grassia, giuseppe.mangioni\}@unict.it}}

\maketitle              

\begin{abstract}
Graph Neural Networks (GNNs) have been widely used to learn representations on graphs and tackle many real-world problems from a wide range of domains.
In this paper we propose \emph{wsGAT}, an extension of the Graph Attention Network (GAT)~\cite{gat2018} layers, meant to address the lack of GNNs that can handle graphs with signed and weighted links, which are ubiquitous, for instance, in trust and correlation networks.
We first evaluate the performance of our proposal by comparing against GCNII~\cite{chen2020simple} in the weighed link prediction task, and against SGCN~\cite{derr2018signed} in the link sign prediction task.
After that, we combine the two tasks and show their performance on predicting the signed weight of links, and their existence.
Our results on real-world networks show that models with \emph{wsGAT} layers outperform the ones with GCNII and SGCN layers, and that there is no loss in performance when signed weights are predicted.
\keywords{network science, link prediction, geometric deep learning, geometric representation learning, graph neural networks}
\end{abstract}
\input{introduction}

\input{formulation}

\input{results}

\input{conclusions}

%
%
\bibliographystyle{splncs03}
\bibliography{article}

\end{document}

%% file: introduction.tex
\section{Introduction}
\label{s:introduction}

Graphs (or networks) are a very flexible formalism that can be used to represent many real-world systems made up of interacting entities~\cite{newman2003structure}.
In a network, nodes represent entities and links model how they interact.
For instance, in a social network, nodes can be people while links among them represent some kind of social interaction, such as friendship or acquaintance relations~\cite{borgatti2018analyzing}.
Depending on the modeled system, links can be also directed~\cite{bang2008digraphs} and/or weighted~\cite{newman2004analysis}.
Directionality means there is an asymmetric relation among nodes, as it happens in a social media, like Twitter, where a user $Bob$ may follow $Alice$, while $Alice$ is not required to follow $Bob$'s Twitter account. 
Moreover, relations can be characterized by different strength levels, that are represented in a network by labelling links with weights expressed in an appropriate scale.

Signed networks~\cite{leskovec2010signed} are another class of networks, where links can be positive or negative.
They are especially used to model good or bad relations among nodes.
A notable example are trust networks~\cite{bachi2012classifying,carchiolo2012trust,carchiolo2013users}, where nodes are users and positive/negative links among them are used to model trust/distrust relations.
In general, signed networks can also be weighted, so relations can be positive or negative and with a given strength.
Considering again trust networks, each link can express more or less strong relationships of trust or distrust.
Weighted and signed networks are also commonly used to represent correlations networks~\cite{mizuno2006correlation,chen2008revealing}, where links among entities express the level of correlation that, in general, can be a positive or negative real number.

To solve many (hard) problems on networks, deep learn techniques have recently been used~\cite{zhou2020graph,wu2020comprehensive,grassia2019learning,grassia2021machine}.
In particular, they employed Graph Neural Networks (GNNs)~\cite{scarselli2008graph} to learn representations on graphs by abstracting from the specific application domain.
GNNs are powerful tools and their applicability has been successfully demonstrated even to solve very complex problems, such as link prediction in complex networks.
Among different GNN layer model, the Graph Attention Networks (GATs)~\cite{gat2018} are one of the most promising, both in terms of performance and flexibility in solving problems in different domains.
The original GAT formulation only took into account (un)directed un-weighted networks.
In this paper we propose \emph{wsGAT}, an extension of the GAT to cope with signed and weighted networks.
We show wsGAT applicability to real-world signed and weighted networks by solving the link prediction task.

We compare wsGAT performance by solving the same task with GCNII~\cite{chen2020simple} and SGCN~\cite{derr2018signed} models, respectively used to perform weighted and signed link prediction.
Our results show that models with wsGAT layers outperform the ones with GCNII and SGCN layers.

The paper is organized as in the following. 
Section~\ref{s:formulation} formally introduces our proposed model and  how  it  works,  whereas  Section~\ref{s:results} illustrates the  experiments carried out on real--world datasets together with a comparison with other approaches.
Finally, Section~\ref{s:conclusions} provides  some  concluding  remarks  and  ideas  for future developments.

%% file: formulation.tex
\section{Formulation}
\label{s:formulation}

In this paper, we extend the \emph{Graph Attention Networks} (GAT)~\cite{gat2018} by modifying the computation of the attention coefficient to also account for the (signed) link weight.

As common in the literature, we indicate with $\mathbf{h}^{(k)}_{n} \in \mathbb{R}^{F_k}$ the node embedding of node $i$ after the $k$-th GNN layer, where $F_k$ is the number of features. 
According to this notation, $\mathbf{h}^{(0)}_{n}$ are the node's input features $\mathbf{x}_n$.

In the original GAT formulation, the authors borrow the \emph{attention} mechanism~\cite{vaswani2017attention}, defined to handle variable length sequences and used successfully in the Natural Language Processing (NLP) field, to assign a (relative) importance score to each of the neighbors of the target node.
Specifically, they compute the \emph{attention}  coefficient $\alpha_{ij}$ of a node $i$ for each neighboring node $j$ as in equation~\ref{e:original_attention}.

\begin{equation}
    \alpha^{k}_{ij} = \mathrm{softmax}(\mathrm{LeakyReLU}(\mathbf{e}^{k}_i))_{j}
    \label{e:original_attention}
\end{equation}

\begin{equation}
    e^{k}_{ij} = \mathbf{a}^{\intercal}_k( \mathbf{W}^{k} \mathbf{h}_i \mathbin\Vert \mathbf{W}^{k} \mathbf{h}_j)
\end{equation}

where $\mathbf{W}^{k} \in \mathbb{R}^{F_k \times F_{k+1}}$ is a learned weight matrix, $\mathbf{a}^{\intercal}_k \in \mathbb{R}^{2 \cdot F_{k+1}}$ is a learned weight vector and $\Vert$ is the concatenation operator.

The attention coefficient is then used to scale the incoming node embedding of the neighbours as in equation~\ref{e:embedding_sum}.

\begin{equation}
    \mathbf{h}_{i}^{(k+1)} = f( \sum_{j \in \mathcal{N}_i \cup \{i\}}  \alpha^{k}_{ij} \mathbf{h}_{j}^{(k)} )
    \label{e:embedding_sum}
\end{equation}

where $f$ is an activation function, $\mathcal{N}_i$ is the neighbourhood of node $i$, and may include the node $i$ itself if self-loops are added to the network.

The main limitation of this formulation is that the same weight matrix $\mathbf{W}^{k}$ is applied independently to both of the embeddings of the target and neighbouring nodes, i.e., they are combined linearly.
To achieve better attention scores, other approaches that use Multi-Layer Perceptrons have been proposed~\cite{grl_book}.

We follow this trend and also account for the (signed) link weight $w_{ij}$ in the attention computation.
In detail, we first modify the computation of $e^{k}_{ij}$ as follows:

\begin{equation}
    e^{k}_{ij} = \mathrm{MLP}^k(\mathbf{h}^{(k)}_i \mathbin\Vert \mathbf{h}^{(k)}_j \mathbin\Vert w_{ij})
\end{equation}

where $MLP^k$ is a Multi-Layer Perceptron with the only requirement that the last layer can also produce negative values (e.g., a zero-centred activation function is used) and $\mathbf{w}_{ij}$ is the weight of the link.

The attention coefficients are then computed as:

\begin{equation}
    \alpha^{k}_{ij} = \mathrm{sign}(e_{ij}) \cdot \mathrm{softmax}(\mathrm{abs}(\mathbf{e}^{k}_i))_{j}
\end{equation}

That is, in our formulation $\alpha_{ij} \in [-1, 1]$, meaning that the contribution of each neighbouring node to equation~\ref{e:embedding_sum} can be positive or negative.


The choice of a Multi-Layer Perceptron allows the network to learn the relative importance of the features of the neighbouring nodes $j$, with respect to the ones of the target node $i$, and is also affected by the weight and sign of the link between them.

As in the original GAT formulation, \emph{wsGAT} also support multi-head attention, meaning that multiple embeddings for a node can be computed --- each using a different set of parameters ---  and concatenated/sum together.

%% file: results.tex
\section{Results}
\label{s:results}

To validate the proposed \emph{wsGAT} layer, we test it in the link and weight prediction task on real-world trust networks.

Since, to the best of our knowledge, no other GNN layer can handle both signed and weighted links, we first decompose the final task of signed and weighted link prediction in two sub-tasks and compare our proposal against the state-of-the-art layers.
Specifically, we first compare on the link sign prediction with \emph{Signed Graph Convolutional} (SGCN), and on the (unsigned) link weight prediction with \emph{GCNII}.

\subsection{Dataset}
We test our proposal on 4 real-world trust networks.
More in detail, we test on the who-trusts-whom networks from the Advogato online community, where trust 4 trust levels can be assigned (corresponding weights are from 0.4 to 1.0 with 0.2 step), from the Bitcoin Alpha and OTC platforms, where scores are on a scale of -10 (total distrust) to +10 (total trust), and from the Epinions.com community, where users can assign a positive or negative trust score to each other. 
We summarize the networks used for the experiments in \tablename~\ref{t:networks}.


\begin{table*}[ht!]
\begin{tabular}{llllllll}
\cline{1-7}
\multicolumn{1}{|l|}{Network} &
  \multicolumn{1}{c|}{$\vert$V$\vert$} &
  \multicolumn{1}{c|}{$\vert$E$\vert$} &
  \multicolumn{1}{c|}{Positive Links} &
  \multicolumn{1}{c|}{Min. Link Weight} &
  \multicolumn{1}{c|}{Max. Link Weight} &
  \multicolumn{1}{c|}{Refs} &
   \\ \cline{1-7}
\multicolumn{1}{|l|}{advogato} & \multicolumn{1}{r|}{6,541} & \multicolumn{1}{r|}{51,127} & \multicolumn{1}{c|}{100\%} & \multicolumn{1}{c|}{0} & \multicolumn{1}{c|}{1} & \multicolumn{1}{c|}{\cite{5380625}} &  \\ \cline{1-7}
\multicolumn{1}{|l|}{bitcoin-alpha} & \multicolumn{1}{r|}{3,783} & \multicolumn{1}{r|}{24,186} & \multicolumn{1}{c|}{89.98\%} & \multicolumn{1}{c|}{-10} & \multicolumn{1}{c|}{10} & \multicolumn{1}{c|}{\cite{7837846}} &  \\ \cline{1-7}
\multicolumn{1}{|l|}{bitcoin-otc} & \multicolumn{1}{r|}{5,881} & \multicolumn{1}{r|}{35,592} & \multicolumn{1}{c|}{93.64\%} & \multicolumn{1}{c|}{-10} & \multicolumn{1}{c|}{10} & \multicolumn{1}{c|}{\cite{7837846}} &  \\ \cline{1-7}
\multicolumn{1}{|l|}{epinions} & \multicolumn{1}{r|}{131,828} & \multicolumn{1}{r|}{841,372} & \multicolumn{1}{c|}{85.29\%} & \multicolumn{1}{c|}{-1} & \multicolumn{1}{c|}{1} & \multicolumn{1}{c|}{\cite{konect:massa05}} &  \\ \cline{1-7}
                                                    &                       &                       &                       & 
\end{tabular}
\caption{\textbf{Dataset.} Details about the networks used in this paper.}
\label{t:networks}
\end{table*}

\subsection{Sign prediction}
In the first sub-task, we perform sign prediction --- i.e. prediction of the kind of relationship (positive or negative) between two nodes in trust networks --- and compare against \emph{SGCN}~\cite{derr2018signed}.

The \emph{SGCN} layer, to the best of our knowledge, is the only one able to handle signed links.
In particular, they use balance theory and compute two feature sets for each node by splitting the node neighbourhood into two sub-neighbourhoods (i.e., one with all the positive links and the other with all the negative ones).
That is, each node has a positive and a negative feature sets.
This is a limitation from a Network Science perspective, as the two sub-networks may have different characteristics w.r.t. the original network, or disconnected components may emerge (e.g., in the case of unbalanced link signs).
However, authors mitigate this issue by influencing each feature set with the other: when computing the positive node features, they also sum a function of the negative ones, and vice-versa.
Another limitation of \emph{SGCNs} is that they do not support link weights, which is useful in many contexts, like the trust one.

For a fair comparison with this approach, we use the same input spectral features and the same train methodology proposed in their paper.
In detail, we provide the Signed Spectral Embedding (SSE) from~\cite{doi:10.1137/1.9781611972801.49} as input node features, and use a node classifier that predicts whether, given a pair of node embeddings, the link between the two nodes is positive, negative or non-existent.
During the training phase, we provide $80\%$ of existing links as train examples (and remove the remaining ones from the network), plus the same number of non-existing ones sampled randomly.
However, in analogy to their methodology, we predict only the sign of existent links during testing.

We employ the Area Under the Curve (AUC) of the Receiver Operating Characteristic (ROC) curve and the F1-scores to evaluate the prediction performance.
According to sign prediction results, shown in \tablename~\ref{t:sign_results}, \emph{wsGAT} outperform the best \emph{SGCN} algorithm on the three signed networks in our dataset.
It is worth noting that the \emph{SGCN} results for the \emph{epinions} network differ from the ones reported by the authors in their paper as we do not filter low-degree nodes from the graph.

{\renewcommand{\arraystretch}{1.2}
\setlength{\tabcolsep}{0.5em}
\begin{table*}[ht!]
\centering
\begin{tabular}{|l|r|r|r|r|r|r|}
\hline
GNN Layer & \multicolumn{2}{c|}{bitcoin-alpha} & \multicolumn{2}{c|}{bitcoin-trust} & \multicolumn{2}{c|}{epinions}   \\ \hline
SGCN2     & 0.796            & 0.917           & 0.823            & 0.925           & \textbf{0.842} & 0.946          \\ \hline
wsGAT     & \textbf{0.832}   & \textbf{0.967}  & \textbf{0.845}   & \textbf{0.953}  & 0.839          & \textbf{0.949} \\ \hline
\end{tabular}
\caption{\textbf{Sign prediction results (ROC AUC | F1).}}
\label{t:sign_results}
\end{table*}
}


\subsection{Weight prediction}
In the second sub-task, we perform link weight prediction -- i.e., predict the (unsigned) strength of the relationship between two nodes ---.
Here, we compare against \emph{GCNII}~\cite{chen2020simple} that were proposed to simplify and improve the Graph Convolutional Networks (GCN) by Kipf et al.~\cite{kipf2017semisupervised}.
For both \emph{wsGAT} and \emph{GCNII} we employ the same model architecture: after the GNN layers we use two Multi-Layer Perceptrons; while both take a pair of node embeddings as input, one is trained to predict if the existence of the link between the input nodes, the other is trained to predict the weight.
Both MLPs in our tests have fixed number of layers (3) and neurons (100 neurons per layer, 1 output).

Regarding the training, we split the network links into training links ($80\%$) and test links ($20\%$, removed before the training).
In addition, for each set we sample the same number of non-existing links to provide the negative examples, and assign a $0$ weight to them.

This time we use the ROC AUC and the F1 scores to evaluate the link prediction performance, and we measure the error on the weight prediction (only for existing links) with the Mean Absolute Error (MAE).
The weight prediction results are reported in \tablename~\ref{t:weight_results}.
\emph{wsGAT} outperform the \emph{GCNII} not only in the link prediction task, but also predict more accurate link weights.



{\renewcommand{\arraystretch}{1.2}
\setlength{\tabcolsep}{0.4em}
\begin{table*}[ht!]
\centering
\begin{tabular}{|l|r|r|r|r|r|r|r|r|r|}
\hline
GNN Layer & \multicolumn{3}{c|}{advogato}               & \multicolumn{3}{c|}{bitcoin-alpha}               & \multicolumn{3}{c|}{bitcoin-trust}                \\ \hline
GCNII     & 0.880          & 0.824          & 0.158 & 0.912          & 0.841          & 0.1470         & 0.909          & 0.830          & 0.179            \\ \hline
wsGAT     & \textbf{0.910} & \textbf{0.839} & \textbf{0.142} & \textbf{0.923} & \textbf{0.851} & \textbf{0.130} & \textbf{0.929} & \textbf{0.860} & \textbf{0.154} \\ \hline
\end{tabular}
\caption{\textbf{Absolute weight prediction results (ROC AUC | F1 | MAE).} Note that while the higher the AUC and F1 scores the better, MAE is an error score and lower values represent smaller errors. }
\label{t:weight_results}
\end{table*}
}

\subsection{Signed weight prediction}
Finally, we merge the two sub-tasks discussed previously and predict the existence of links and of their signed weight.

As in the weight prediction sub-task, we use the AUC and the F1 to measure the link prediction performance, and the MAE to measure the error on the weight prediction.

The results on the signed and weighted Bitcoin networks, reported in \tablename~\ref{t:signed_weight_results}, show that the link prediction performance is almost the same as the un-signed case, and the mean absolute error (now on a scale from -10 to +10) drops significantly.

{\renewcommand{\arraystretch}{1.2}
\setlength{\tabcolsep}{0.5em}
\begin{table*}[ht!]
\centering
\begin{tabular}{|l|r|r|r|r|r|r|}
\hline
GNN Layer & \multicolumn{3}{c|}{bitcoin-alpha} & \multicolumn{3}{c|}{bitcoin-trust} \\ \hline
wsGAT     & 0.922     & 0.839     & 0.069    & 0.921     & 0.852    & 0.079    \\ \hline
\end{tabular}
\caption{\textbf{Signed weight prediction results (ROC AUC | F1 | MAE).} Note that while the higher the AUC and F1 scores the better, MAE is an error score and lower values represent smaller errors.}
\label{t:signed_weight_results}
\end{table*}
}

\subsection{Code availability}
\emph{wsGAT} were implemented on top of PyTorch Geometric~\cite{Fey/Lenssen/2019} v1.6.3.
Code will be publicly available after the publication of the paper at the following URL https://github.com/NetworkScienceLab/wsGAT.

%% file: conclusions.tex
\section{Conclusions}
\label{s:conclusions}

In this paper we present \emph{wsGAT}, an extension of \emph{Graph Attention Networks} (GAT) to signed and weighted networks.
The results in link prediction tasks on signed and weighted real-world trust networks and the comparison with state-of-the-art algorithms confirm the validity of our approach, that provides a useful tool for many research and application scenarios, not limited to link prediction. 